\documentclass[english,nofootinbib, notitlepage, pra, english]{revtex4-1}
\usepackage[T1]{fontenc}
\usepackage[utf8]{inputenc}
\usepackage{geometry}
\usepackage{color}
\usepackage{babel}
\usepackage{array}
\usepackage{multirow}
\usepackage{amsmath}
\usepackage{graphicx}
\usepackage[unicode=true,pdfusetitle,
 bookmarks=true,bookmarksnumbered=true,bookmarksopen=true,bookmarksopenlevel=1,
 breaklinks=false,pdfborder={0 0 0},pdfborderstyle={},backref=false,colorlinks=true]
 {hyperref}

\makeatletter

\providecommand{\tabularnewline}{\\}

\usepackage[section]{placeins}

\makeatother

\begin{document}
\global\long\def\seventh#1#2#3#4#5#6#7{\left[#1\cdot\left(#2\times#3\right)\right]\left(#4\cdot#5\right)\left(#6\cdot#7\right)}%

\title{Disentangling enantiosensitivity from dichroism using bichromatic
fields}

\author{Andres F. Ordonez}
\email{ordonez@mbi-berlin.de}
\affiliation{Max-Born-Institut, Berlin, Germany}
\affiliation{Technische Universit\"at Berlin, Berlin, Germany}
\author{Olga Smirnova}
\email{smirnova@mbi-berlin.de}
\affiliation{Max-Born-Institut, Berlin, Germany}
\affiliation{Technische Universit\"at Berlin, Berlin, Germany}

\begin{abstract}
We discuss how tensorial observables, available in photoelectron angular
distributions resulting from interaction between isotropic chiral
samples and cross polarized $\omega$-$2\omega$ bichromatic fields,
allow for chiral discrimination without chiral light and within the
electric-dipole approximation. We extend the concept of chiral setup
\href{https://link.aps.org/doi/10.1103/PhysRevA.98.063428}{{[}Phys. Rev. A 98, 063428 (2018){]}}, which explains how chiral discrimination
can be achieved in the absence of chiral light, to the case of tensorial
observables. We derive selection rules for the enantiosensitivity
and dichroism of the $b_{l,m}$ coefficients describing the photoelectron
angular distribution valid for both weak and strong fields and for
arbitrary $\omega$-$2\omega$ relative phase. Explicit expressions
for simple perturbative cases are given. We find that, besides the
dichroic non-enantiosensitive \href{https://aip.scitation.org/doi/10.1063/1.5111362}{{[}J. Chem. Phys. 151 074106 (2019){]}},
and dichroic-and-enantiosensitive $b_{l,m}$ coefficients found recently
\href{https://link.aps.org/doi/10.1103/PhysRevA.99.063406}{{[}Phys. Rev. A 99, 063406 (2019){]}}, there are also enantiosensitive
non-dichroic $b_{l,m}$ coefficients. These reveal the molecular enantiomer
independently of the relative phase between the two colors and are
therefore observable even in the absence of stabilization of the $\omega$-$2\omega$
relative phase. 
\end{abstract}
\maketitle

\section{Introduction}

More than two centuries after the pioneering observations of Biot
and Arago \citep{lowry_optical_1964}, the interaction between light
and chiral matter \citep{berova_comprehensive_2012} remains a very
active field of research \citep{eibenberger_enantiomer-specific_2017,yachmenev_field-induced_2019,goetz_quantum_2019,ayuso_synthetic_2019,cao_lateral_2018}.
This research effort is fueled not only by the interest in finding
new ways of manipulating light and matter but also by the homochirality
of life. While the molecule-molecule interactions that take place
in biological systems are often strongly enantiosensitive \citep{noyori_asymmetric_2002},
the enantiosensitive response in ``traditional'' light-matter interactions
is usually very weak. This weakness, which limits the potential of
light-based applications, is prevalent in situations where the electric-dipole
approximation is well justified but the chiral effects appear through
small corrections such as the magnetic-dipole interaction \citep{barron_molecular_2004}.
Besides ingenious methods to cope with such situations \citep{rhee_femtosecond_2009,tang_enhanced_2011},
weakly enantiosensitive responses can be avoided from the outset by
relying on chiral effects occurring within the electric-dipole approximation
\citep{ordonez_generalized_2018-1,ordonez_chiral_2019}. 

Among the chiral electric-dipole effects, photoelectron circular dichroism
(PECD) \citep{ritchie_theory_1976,bowering_asymmetry_2001,powis_photoelectron_2000}
is very well established and has been shown to consistently yield
strongly enantiosensitive signals across many molecular species \citep{nahon_valence_2015}
and different photoionization regimes \citep{beaulieu_universality_2016}.
In PECD, isotropically oriented chiral molecules are photoionized
using circularly polarized light and the photoelectron angular distribution
displays a so-called forward-backward asymmetry (asymmetry with respect
to the polarization plane). This asymmetry is both enantiosensitive
(opposite for opposite enantiomers) and dichroic (opposite for opposite
polarizations), and results from the lack of mirror symmetry of the
chiral molecules (and thus of the light-matter system) with respect
to the plane of polarization. For an isotropically oriented achiral
molecule and within the electric-dipole approximation (i.e. ignoring
light-propagation effects), the light-matter system is mirror-symmetric
with respect to the polarization plane and therefore forward-backward
asymmetric observables resulting from a single-molecule response are
symmetry-forbidden. For a randomly oriented chiral molecule the mirror
symmetry of the light-matter system is absent and there are usually
no further symmetries preventing the emergence of forward-backward
asymmetric observables. In the one-photon case, the forward-backward
asymmetry results from the non-zero molecular rotational invariant
describing the average circular polarization of the photoionization-dipole
vector field \citep{ordonez_method_PAD}. 

Very recently, the response of chiral molecules to more elaborate
field polarizations has been investigated in the multiphoton and strong-field
regimes \citep{demekhin_photoelectron_2018,demekhin_photoelectron_2019,rozen_controlling_2019}.
In these works, a bichromatic field with frequencies $\omega$ and
$2\omega$ linearly polarized perpendicular to each other yields a
new type of enantiosensitive and dichroic asymmetry in the photoelectron
angular distribution. The Lissajous figure of this bichromatic field
has an eight-like shape for particular phase relations between the
two colors, and therefore ``rotates'' in opposite directions in
the first and second halves of its cycle, with the direction of rotation
of the field locked to the sign of its $\omega$ component. For example,
the rotation is clockwise when the $\omega$ component is positive
and counterclockwise when the $\omega$ component is negative. The
observed asymmetry in the photoelectron angular distribution corresponds
to a correlation between the so-called forward-backward direction
and the up-down direction determined by the $\omega$ field (see Fig.
1 in Ref. \citep{demekhin_photoelectron_2018}). So far, this asymmetry
has been analyzed on the basis of a subcycle PECD-like picture where
electrons detected in the upper hemisphere have an e.g. positive forward-backward
asymmetry because they were produced by a field rotating e.g. clockwise,
while the electrons detected in the lower hemisphere have a negative
forward-backward asymmetry because they were produced by a field rotating
counterclockwise \citep{demekhin_photoelectron_2018,rozen_controlling_2019}.
Extensions of this reasoning to account for non-zero asymmetries for
other relative phases between the $\omega$ and $2\omega$ components
have been considered in Refs. \citep{demekhin_photoelectron_2019}
and \citep{rozen_controlling_2019}.

Here we approach the description of the photoelectron angular distribution
taking explicitly into account its tensorial character, the role of
tensorial detectors in forming the chiral setup required to distinguish
between opposite enantiomers, and the symmetry of the light-matter
system.  In Sec. \ref{sec:Chiral-probes} we discuss general considerations
regarding how to distinguish between opposite enantiomers in isotropic
samples with and without relying on the chirality of a light field.
In Sec. \ref{sec:Tensor-observables-and-chiral-setups} we discuss
how tensorial observables can be used for the construction of chiral
setups. In section \ref{sec:cross-polarized} we consider the interaction
of an $\omega$-2$\omega$ cross polarized field with isotropic molecular
samples and derive selection rules for multipolar observables. Then
we consider the perturbative description of photoionization and exploit
the method presented in Ref. \citep{ordonez_method_PAD} to obtain
explicit formulas for some representative $\tilde{b}_{l,m}$ coefficients
of the photoelectron angular distribution displaying different combinations
of dichroism and enantiosensitivity. Finally, we list our conclusions
in Sec. \ref{sec:Conclusions}. Analogous results but for the case
of charge multipoles induced via excitation of bound states is presented
in Ref. \citep{ordonez_inducing_2020}. 

\section{Chiral probes\label{sec:Chiral-probes}}

Distinguishing between the left version (L) and the right version
(R) of a chiral object invariably requires interaction with another
chiral object (say R'). The difference between the interactions L+R'
and R+R' is the essence of any enantiosensitive phenomenon. Here we
are interested in the phenomena where light is used to distinguish
between opposite enantiomers of an isotropically oriented chiral molecule.
In the simplest case, one lets circularly polarized light of a given
handedness interact with a chiral sample and measures a scalar, namely
the amount of light that is absorbed. In this situation, known as
circular dichroism (CD), the chiral probe is the circularly polarized
light and its handedness (a pseudoscalar) is given by its helicity,
i.e. the projection of the photon spin (the rotation direction of
the light at a given point in space, a pseudovector) on the propagation
direction of the light (a vector). Another canonical example is optical
activity, where one passes linearly polarized light through a chiral
sample and measures the rotation of the polarization plane. In this
case one measures an angle (a pseudovector). To measure it one must
define a positive and a negative direction in the laboratory frame.
Although such definition is just as arbitrary as defining what is
left and what is right, it is a fundamental step in the measurement
process. Once it has been defined, the handedness of the probe is
given by the projection of the positive unit angle pseudovector on
the propagation direction of the light. That is, the chiral probe
in this case is the chiral setup formed by the (achiral) light and
the (achiral) detector (which encodes the definition of positive and
negative rotations) together. This shows how a chiral setup may probe
the handedness of a molecule in the absence of chiral light \citep{ordonez_chiral_2019}.

Circular dichroism and optical activity have in common that the handedness
of the chiral probe relies on the propagation direction of the light.
Since that propagation direction is immaterial within the electric-dipole
approximation, unless one considers corrections to the electric-dipole
approximation the probe ceases to be chiral and both effects vanish.
To the extent that such corrections are typically small at the single
molecule level, these effects are also correspondingly small. In order
to obtain bigger enantiosensitive signals at the single-molecule level,
a probe which is chiral within the electric-dipole approximation is
required. Furthermore, if the result of the measurement is a scalar
(like in circular dichroism), the light itself must be chiral. The
concept of light which is chiral within the electric-dipole approximation,
i.e. locally-chiral light, has been recently developed in Ref. \citep{ayuso_synthetic_2019}.
If the result of the measurement is a polar vector, such as a photoelectron
current, then the detector required to measure the vector must define
(again, in an arbitrary fashion) a positive and a negative direction.
In addition, if the polarization of the light allows the definition
of a pseudovector, as is the case for example for circularly polarized
light, where the pseudovector indicates the photon's spin, then one
can define a chiral setup with its handedness given by the projection
of the light pseudovector on the positive direction defined by the
detector. This type of chiral setup is common to a series of recently
discovered phenomena that range from rotational dynamics \citep{patterson_enantiomer-specific_2013,patterson_sensitive_2013,lehmann_chapter_2018}
to photoionization \citep{ritchie_theory_1976,bowering_asymmetry_2001,powis_photoelectron_2000,lux_circular_2012},
and it was recently discussed in Ref. \citep{ordonez_generalized_2018-1}.
In what follows we will extend the concept of chiral setups, to include
those that rely on tensors of rank 2 (relevant for the results in
Refs. \citep{demekhin_photoelectron_2018,demekhin_photoelectron_2019,rozen_controlling_2019})
and higher.

\section{Tensor observables and chiral setups\label{sec:Tensor-observables-and-chiral-setups}}

\begin{figure}
\begin{centering}
\includegraphics[scale=0.2]{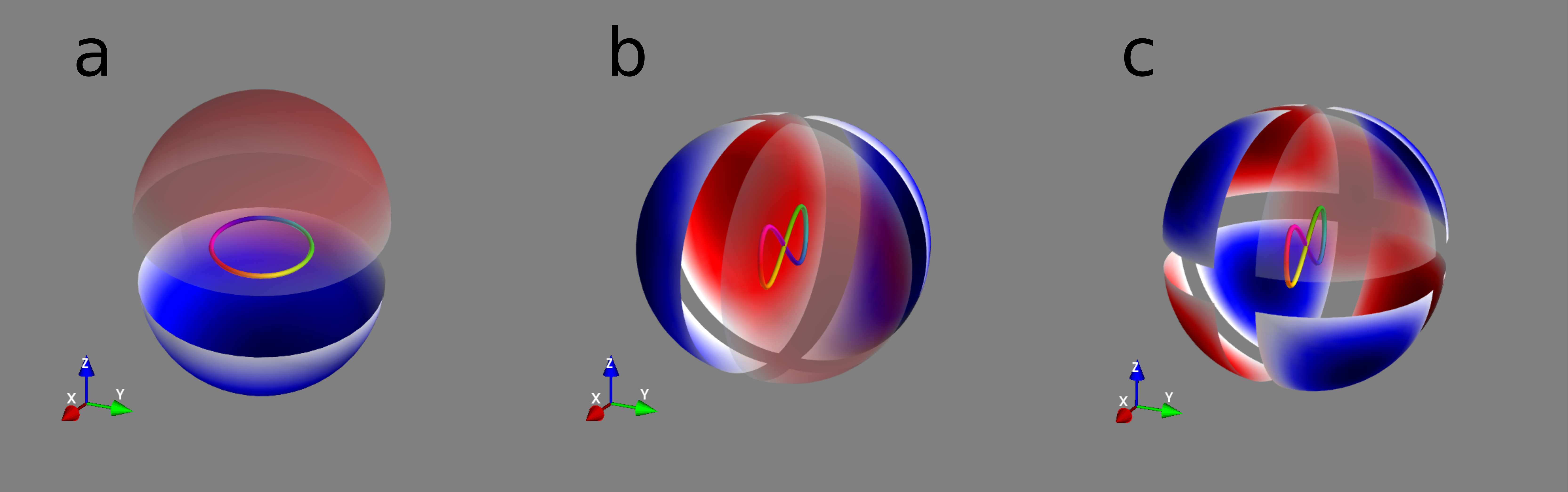}
\par\end{centering}
\caption{Chiral setups consisting of a Lissajous figure and detectors. Counts
on red/blue detectors are added/subtracted. The color of the Lissajous
figure indicates the phase of the oscillation. (a) A field circularly
polarized in the $xy$ plane and a detector for $\tilde{b}_{1,0}$.
(b) and (c) A cross-polarized $\omega$-2$\omega$ field in the $xz$
plane and a detector for $\tilde{b}_{2,-2}$ and $\tilde{b}_{3,-2}$
respectively. \label{fig:Chiral-setups}}
\end{figure}

Second and higher-rank tensors emerge naturally for observables which
depend on a vector. Charge densities and photoelectron angular distributions
(PADs) offer exactly such kind of observable and, as any other function
which depends on a vector, they can be expanded as

\begin{equation}
W(\vec{k})=\sum_{l,m}\tilde{b}_{l,m}\left(k\right)\tilde{Y}_{l}^{m}(\hat{k}),\label{eq:W}
\end{equation}

where $\vec{k}$ is the photoelectron momentum and we have chosen
to do the expansion in terms of real spherical harmonics $\tilde{Y}_{l}^{m}$.
In the case of a charge distribution we replace the photoelectron
momentum $\vec{k}$ by the position $\vec{r}$ and $\tilde{b}_{l,m}$
could be a time dependent quantity (see \citep{ordonez_inducing_2020}).
The $\tilde{b}_{l,m}$ coefficients not only encode all the information
contained in $W(\vec{k})$ but are also examples of tensors of rank
$l$ (see e.g. Sec. 4.10 in Ref. \citep{brink_angular_1968}).

In principle, the measurement of a particular $\tilde{b}_{l,m}$ coefficient
can be performed directly by using a detector with a structure that
reflects the corresponding $\tilde{Y}_{l}^{m}$. For example, a detector
for $\tilde{b}_{0,0}$ would simply sum the counts in all directions.
A detector for $\tilde{b}_{1,0}$ would have two plates arranged as
in Fig. \ref{fig:Chiral-setups}a, so that it would add all the counts
on one of them (red) and subtract all the counts on the other (blue).
The choice of which plate adds and which plate subtracts is what physically
defines the $\hat{z}$ direction of the laboratory frame. A detector
for $\tilde{b}_{2,-2}$ would have four plates arranged as in Fig.
\ref{fig:Chiral-setups}b, with a pair of opposite plates adding counts
and the other pair subtracting counts. In this case, the detector
defines the directions that correspond to a positive correlation between
$x$ and $y$, and those that correspond to a negative correlation
between $x$ and $y$. Analogously, a detector for $\tilde{b}_{3,-2}$
has eight plates arranged as in Fig. \ref{fig:Chiral-setups}c, and
distinguishes positive from negative correlations of $x$, $y$, and
$z$.

If we now take into account the electric field, then the combination
of a $\tilde{b}_{l,m}$-specific detector and the Lissajous figure
of the electric field can make up a chiral setup as shown in Fig.
\ref{fig:Chiral-setups}. These setups are non-superimposable on their
mirror images. This is particularly simple to see in Fig. \ref{fig:Chiral-setups}
for reflections with respect to the polarization plane, which do not
change the Lissajous figure but do swap blue and red plates.

Of course, a chiral setup is only relevant if the light-matter system
is indeed asymmetric enough that it can yield a corresponding non-zero
$\tilde{b}_{l,m}$. In other words, the concept of a chiral setup
answers the question of how we can distinguish opposite enantiomers
without relying on the chirality of light, but it is the lack of symmetry
of the light-matter system itself the deciding factor which determines
the emergence of an enantiosensitive observable in the first place
(see e.g. Fig. 2 in Ref. \citep{ordonez_generalized_2018-1}). We
now turn to the analysis of a specific family of Lissajous figures
to illustrate this in detail.

\section{$\boldsymbol{\omega}$-2$\boldsymbol{\omega}$ cross polarized field\label{sec:cross-polarized}}

\begin{figure}
\begin{centering}
\includegraphics[width=0.8\textwidth]{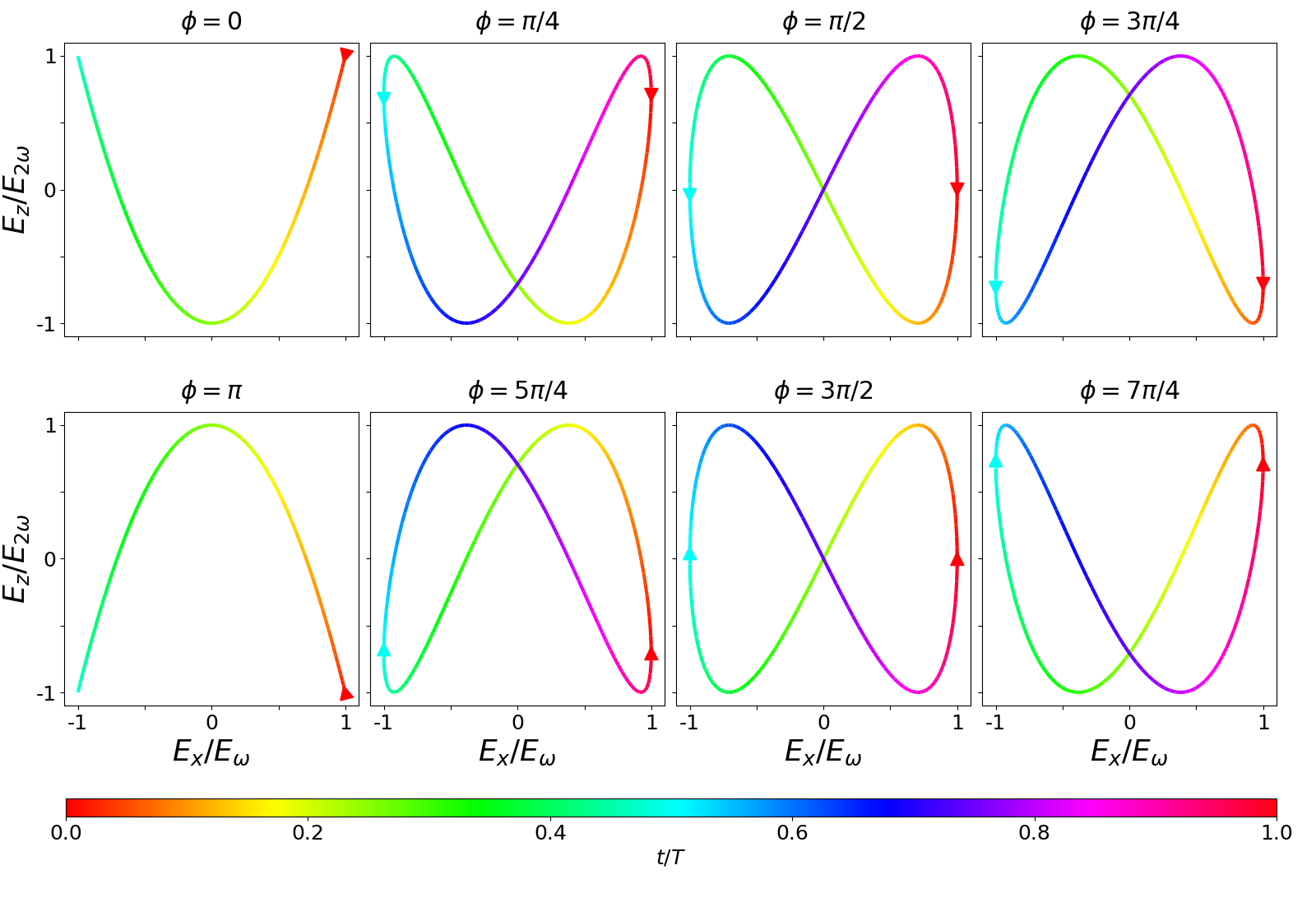}
\par\end{centering}
\caption{Lissajous curves of the field (\ref{fig:Lissajous-curves}) for different
values of $\phi$. The position of the red arrow indicates $t=0$.\label{fig:Lissajous-curves}}
\end{figure}

Consider a two-color field of the form 

\begin{equation}
\vec{E}\left(t\right)=E_{\omega}\cos\left(\omega t\right)\hat{x}+E_{2\omega}\cos\left(2\omega t+\phi\right)\hat{z}\label{eq:field}
\end{equation}

This field is illustrated in Fig. \ref{fig:Lissajous-curves} for
different phases $\phi$\footnote{Our choice of axes, which differs from Refs. \citep{demekhin_photoelectron_2018,demekhin_photoelectron_2019,rozen_controlling_2019},
 simplifies Tab. \ref{tab:conditions}. Our choice of phase coincides
with that in Ref. \citep{rozen_controlling_2019} and differs from
the one in Refs. \citep{demekhin_photoelectron_2018,demekhin_photoelectron_2019}.
Note that some of the plots of Fig. 2(a) in Ref. \citep{rozen_controlling_2019}
are labeled with the wrong phase.}. If we denote rotations by $\pi$ around the $z$ axis by $\hat{R}_{z}^{\pi}$
and time shifts of $\pi/\omega$ by $T_{\pi/\omega}$, then the joint
system (field and isotropic chiral molecules) is invariant with respect
to $\hat{R}_{z}^{\pi}\hat{T}_{\pi/\omega}$. Clearly, the resulting
observables must also be invariant with respect to $\hat{R}_{z}^{\pi}\hat{T}_{\pi/\omega}$.
That is, the symmetry-allowed $\tilde{b}_{l,m}$ coefficients must
satisfy $\hat{R}_{z}^{\pi}\hat{T}_{\pi/\omega}\tilde{b}_{l,m}\tilde{Y}_{l}^{m}=\tilde{b}_{l,m}\tilde{Y}_{l}^{m}$.
This corresponds to two scenarios: either $\hat{R}_{z}^{\pi}\tilde{Y}_{l}^{m}=\tilde{Y}_{l}^{m}$
and $\tilde{b}_{l,m}$ contains only frequencies $2n\omega$, or $\hat{R}_{z}^{\pi}\tilde{Y}_{l}^{m}=-\tilde{Y}_{l}^{m}$
and $\tilde{b}_{l,m}$ contains only frequencies $\left(2n+1\right)\omega$,
where $n=0,1,2,\dots$ Furthermore, since the a reflection $\hat{\sigma}_{y}$
with respect to the $y=0$ plane swaps the enantiomers while leaving
the field invariant, an enantiosensitive $\tilde{b}_{l,m}$ is associated
to a $\tilde{Y}_{l}^{m}$ satisfying $\hat{\sigma}_{y}\tilde{Y}_{l}^{m}=-\tilde{Y}_{l}^{m}$.
And, since a rotation $\hat{R}_{x}^{\pi}$ of $\pi$ around the $x$
axis changes the phase $\phi$ in the field by $\pi$ while leaving
the molecules invariant, a dichroic\emph{}\footnote{Here we use the word dichroic in analogy to how it is used in the
circularly polarized case in PECD, where a change of $\pi$ in the
phase between the two perpendicular components of the field leads
to a change of sign of $b_{1,0}$.} $\tilde{b}_{l,m}$ is associated to a $\tilde{Y}_{l}^{m}$ satisfying
$\hat{R}_{x}^{\pi}\tilde{Y}_{l}^{m}=-\tilde{Y}_{l}^{m}$. These properties,
which are valid independently of the ionization regime, are summarized
in Tabs. \ref{tab:conditions} and \ref{tab:conditions2}. 

Evidently, the values of $l$ and $m$ that determine whether a given
$\tilde{b}_{l,m}$ is symmetry-allowed and whether it is enantiosensitive
and/or dichroic depends on how the field is oriented with respect
to the axes. The choice made here {[}Eq. (\ref{eq:field}){]} yields
the rather simple conditions in Tabs. \ref{tab:conditions} and \ref{tab:conditions2}.
Conditions corresponding to other choices are given in Appendix \ref{subsec:alternative_orientations}
for the sake of comparison with Refs. \citep{demekhin_photoelectron_2018,demekhin_photoelectron_2019,rozen_controlling_2019}. 

Tables \ref{tab:conditions} and \ref{tab:conditions2} reveal two
important features. First, the spatial structure of the response depends
on whether it oscillates with a frequency which is an even (including
zero) or an odd multiple of the fundamental frequency (see also \citep{neufeld_floquet_2019}).
Second, in contrast to a circularly polarized field where dichroism
goes hand in hand with enantiosensitivity, the symmetry of the field
(\ref{eq:field}) leads to four types of signals: 

\renewcommand{\labelenumi}{(\roman{enumi})}
\begin{enumerate}
\item Non-dichroic non-enantiosensitive.
\item Dichroic and enantiosensitive.
\item Dichroic non-enantiosensitive.
\item Enantiosensitive non-dichroic.
\end{enumerate}
Type (i) and type (ii) signals are well known from traditional CD
and PECD. Type (iii) signals are well known in atoms subject to light
fields whose Lissajous figure is not symmetric under spatial inversion
\citep{yin_asymmetric_1992}. They have also been recently calculated
for the case of randomly oriented chiral molecules \citep{goetz_perfect_2019}
for parallel polarizations of $\omega$ and $2\omega$. Here, type
(iii) signals are due to the asymmetry of the field (\ref{eq:field})
along the $z$ direction (see Fig. \ref{fig:Lissajous-curves}). Type
(iv) signals are more exotic but apparently they can also occur in
the context of magnetic effects beyond the electric-dipole approximation
\citep{wagniere_inverse_1989,wagniere_chirality_2011}, where as far
as the authors know they remain to be confirmed by experiment. As
we will show next, photoionization may prove to be a better candidate
for experimentally measuring type (iv) signals.

\begin{center}
\begin{table}
\begin{centering}
\par\end{centering}
\begin{centering}
\par\end{centering}
\begin{centering}
\par\end{centering}
\begin{centering}
\par\end{centering}
\begin{centering}
\begin{tabular}{cc}
\hline 
$\tilde{b}_{l,m}$ & $\left(l,m\right)$ condition\tabularnewline
\hline 
Symmetry-allowed at $2n\omega$ & even $m$\tabularnewline
Symmetry-allowed at $\left(2n+1\right)\omega$ & odd $m$\tabularnewline
Enantiosensitive & $m<0$\tabularnewline
\multirow{2}{*}{Dichroic} & {[}(odd $l-m$) and ($m\geq0$){]} or \tabularnewline
 & (even $l-m$) and ($m<0$)\tabularnewline
\end{tabular}
\par\end{centering}
\caption{Conditions imposed by symmetry on the $\tilde{b}_{l,m}$ coefficients
for a field (\ref{eq:field}) interacting with an isotropic sample
of molecules.\label{tab:conditions}}
\end{table}
\par\end{center}

\begin{table}

\begin{centering}
\par\end{centering}
\begin{centering}
\par\end{centering}
\begin{centering}
\par\end{centering}

\begin{centering}
\begin{tabular}{ccc}
\hline 
Symmetry-allowed $\tilde{b}_{l,m}$ & $\left(l,m\right)$ condition at $2n\omega$ & $\left(l,m\right)$ condition at $\left(2n+1\right)\omega$\tabularnewline
\hline 
Enantiosensitive and dichroic & (even $l$) and (even $m<0$) & (odd $l$) and (odd $m<0$)\tabularnewline
Enantiosensitive non-dichroic & (odd $l$) and (even $m<0$) & (even $l$) and (odd $m<0$)\tabularnewline
Dichroic non-enantiosensitive & (odd $l$) and (even $m>0$) & (even $l$) and (odd $m>0$)\tabularnewline
\hline 
\end{tabular}
\par\end{centering}
\caption{$\left(l,m\right)$ conditions derived from Tab. \ref{tab:conditions}
for non-vanishing enantiosensitive and/or dichroic $\tilde{b}_{l,m}$
coefficients.  \label{tab:conditions2}}
\end{table}

\subsubsection{Photoionization\label{subsec:Photoionization}}

The photoelectron angular distribution accumulated over many cycles
of the field (\ref{eq:field}) corresponds to a signal with zero frequency
and must satisfy the conditions given in Tab. \ref{tab:conditions}
for frequencies $2n\omega$. For convenience, we list the properties
of the symmetry-allowed $\tilde{b}_{l,m}$ coefficients for $l$ up
to four in Tab. \ref{tab:blm_symmetry}. From this table we can see
that e.g. $\tilde{b}_{1,0}$ is dichroic but not enantiosensitive,
$\tilde{b}_{2,-2}$ is enantiosensitive and dichroic, and $\tilde{b}_{3,-2}$
is enantiosensitive but not dichroic. Since $\tilde{Y}_{1}^{0}(\hat{k})\propto k_{z}$,
$\tilde{Y}_{2}^{-2}(\hat{k})\propto k_{x}k_{y}$, and $\tilde{Y}_{3}^{-2}(\hat{k})\propto k_{x}k_{y}k_{z}$,
then $\tilde{b}_{1,0}$ is associated to asymmetry along the direction
of the $2\omega$ field ($\hat{z}$), $\tilde{b}_{2,-2}$ is associated
to correlations between the direction of the $\omega$ field ($\hat{x}$)
and the direction perpendicular to the polarization plane ($\hat{y}$),
and $\tilde{b}_{3,-2}$ is associated to correlations of the three
directions corresponding to the $\omega$ field ($\hat{x}$), the
$2\omega$ field $(\hat{z})$, and the perpendicular to the polarization
plane ($\hat{y}$).

\begin{figure}
\begin{centering}
\includegraphics[width=0.25\paperwidth]{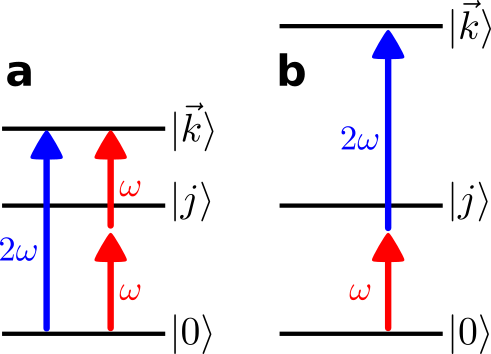}
\par\end{centering}
\caption{\textbf{a.} Simplest scheme to generate a non-zero $\tilde{b}_{1,0}$
(dichroic non-enantiosensitive) and $\tilde{b}_{2,-2}$ (dichroic
and enantiosensitive) using the field (\ref{eq:field}). \textbf{b.}
Simplest scheme to generate a non-zero $\tilde{b}_{3,-2}$ (enantiosensitive
non-dichroic). $\vert0\rangle$ and $\vert j\rangle$ are bound states
and $\vert\vec{k}\rangle$ is a continuum state. \label{fig:energy-scheme}}
\end{figure}

In Ref. \citep{ordonez_method_PAD} we showed that, in general, a
$\tilde{b}_{l,m}$ coefficient is enantiosensitive if and only if
it results from interference between pathways with $N_{1}$ and $N_{2}$
photons, respectively, and $l+N_{1}+N_{2}$ is odd; or if it results
from a direct pathway and $l$ is odd. These conditions together with
Tab. \eqref{tab:conditions2} tell us that a dichroic non-enantiosensitive
$\tilde{b}_{l,m}$ (with odd $l$) can only occur as the result of
interference between pathways involving an even and an odd number
of photons, respectively (so that $N_{1}+N_{2}$ is odd and $l+N_{1}+N_{2}$
is even). For example, $\tilde{b}_{1,0}$ contributes to the photoelectron
peak where absorption of two $\omega$ photons interferes with absorption
of one $2\omega$ photon (see Fig. \ref{fig:energy-scheme}a). The
same condition (odd $N_{1}+N_{2}$) applies for a dichroic and enantiosensitive
$\tilde{b}_{l,m}$ (even $l$) such as $\tilde{b}_{2,-2}$. In contrast,
an enantiosensitive non-dichroic $\tilde{b}_{l,m}$ (odd $l$) can
occur as the result of either a direct pathway involving at least
one $\omega$ photon and one $2\omega$ photon, or as the result of
interference between two pathways both with an even or both with an
odd number of photons (even $N_{1}+N_{2}$). For example, $\tilde{b}_{3,-2}$
contributes to the photoelectron peak corresponding to absorption
of one $\omega$ photon followed by absorption of one $2\omega$ photon
(see Fig. \ref{fig:energy-scheme}b) and vice versa, or as the result
of interference between absorption of two $2\omega$ photons and four
$\omega$ photons. 
\begin{center}
\begin{table}
\begin{centering}
\par\end{centering}
\begin{centering}
\par\end{centering}
\begin{centering}
\par\end{centering}
\begin{centering}
\par\end{centering}
\begin{centering}
\par\end{centering}
\begin{centering}
\begin{tabular}{cccccccccccccc}
\hline 
Symmetry-allowed & $\tilde{b}_{0,0}$ & $\tilde{b}_{1,0}$ & $\tilde{b}_{2,-2}$ & $\tilde{b}_{2,0}$ & $\tilde{b}_{2,2}$ & $\tilde{b}_{3,-2}$ & $\tilde{b}_{3,0}$ & $\tilde{b}_{3,2}$ & $\tilde{b}_{4,-4}$ & $\tilde{b}_{4,-2}$ & $\tilde{b}_{4,0}$ & $\tilde{b}_{4,2}$ & $\tilde{b}_{4,4}$\tabularnewline
\hline 
Enantiosensitive & N & N & Y & N & N & Y & N & N & Y & Y & N & N & N\tabularnewline
Dichroic & N & Y & Y & N & N & N & Y & Y & Y & Y & N & N & N\tabularnewline
\hline 
\end{tabular}
\par\end{centering}
\caption{Dichroism and enantiosensitivity of symmetry-allowed $\tilde{b}_{l,m}$
coefficients describing the photoelectron angular distribution (\ref{eq:W})
resulting from interaction with the field (\ref{eq:field}) for $l$
up to 4. N=No and Y=Yes.  \label{tab:blm_symmetry}}
\end{table}
\par\end{center}

As an example, let us calculate explicit expressions for $\tilde{b}_{1,0}$,
$\tilde{b}_{2,-2}$, and $\tilde{b}_{3,-2}$.  For the process depicted
in Fig. \ref{fig:energy-scheme}a , the interference between the
two pathways  gives rise to a non-zero orientation-averaged $\tilde{b}_{1,0}$
given by \citep{ordonez_method_PAD} (see Appendix)

\begin{equation}
\tilde{b}_{1,0}\left(k\right)=A^{\left(1\right)*}A^{\left(2\right)}g_{1,0}f_{1,0}+\mathrm{c.c.},\label{eq:b_10}
\end{equation}

where $A^{\left(1\right)}$ and $A^{\left(2\right)}$ are complex-valued
constants depending on detunings and pulse envelopes, c.c. denotes
the complex conjugate, $g_{1,0}$ is a (complex-valued) molecular
rotational invariant, and $f_{1,0}$ is a setup (i.e. field + detector)
rotational invariant. $g_{1,0}$ is a scalar\footnote{Explicit expressions for the molecular rotational invariants $g_{1,-1}$,
$g_{2,1}$, and $g_{3,-2}$ are given in Appendix \ref{subsec:appendix_blm_photoionization}.\label{fn:expressions_g}} (in contrast to a pseudoscalar) and therefore $\tilde{b}_{1,0}$
is not enantiosensitive. The expression for $f_{1,0}$ reads as

\begin{equation}
f_{1,0}\equiv\left(\hat{z}\cdot\vec{E}_{2\omega}^{*}\right)\left(\vec{E}_{\omega}\cdot\vec{E}_{\omega}\right)=E_{2\omega}E_{\omega}^{2}e^{i\phi},\label{eq:f_10}
\end{equation}

where $\vec{E}_{2\omega}\equiv E_{2\omega}e^{-i\phi}\hat{y}$, $\vec{E}_{\omega}\equiv E_{\omega}\hat{z}$.
 Equation (\ref{eq:f_10}) shows that the setup rotational invariant
$f_{1,-1}$ is a scalar involving the field vectors $\vec{E}_{2\omega}$
and $\vec{E}_{\omega}$, and the axis $\hat{z}$. As discussed in
Sec. \ref{sec:Tensor-observables-and-chiral-setups}, the axis vector
$\hat{z}$ is defined by the detector needed to measure $\tilde{b}_{1,0}$.
From Eq. (\ref{eq:f_10}) it is evident that $f_{1,0}$ changes sign
when $\phi$ is shifted by $\pi$, and therefore $\tilde{b}_{1,0}$
is dichroic. 

Similarly, the interference between the two pathways in Fig. \ref{fig:energy-scheme}a
also gives rise to a non-zero orientation-averaged $\tilde{b}_{2,-2}$
given by (see Appendix \ref{subsec:appendix_blm_photoionization})

\begin{equation}
\tilde{b}_{2,-2}=A^{\left(1\right)*}A^{\left(2\right)}g_{2,-2}f_{2,-2}+\mathrm{c.c.},\label{eq:b_2m2}
\end{equation}

where the molecular rotational invariant $g_{2,-2}$ is a (complex-valued)
pseudoscalar\textsuperscript{\ref{fn:expressions_g}} and therefore
$\tilde{b}_{2,-2}$ is enantiosensitive. The setup rotational invariant,
given by

\begin{equation}
f_{2,-2}\equiv\left[\hat{x}\cdot\left(\hat{y}\times\vec{E}_{2\omega}^{*}\right)\right]\left(\vec{E}_{\omega}\cdot\vec{E}_{\omega}\right)=E_{2\omega}E_{\omega}^{2}e^{i\phi},\label{eq:f_2m2}
\end{equation}

is also a pseudoscalar and it involves the $\hat{x}$ and $\hat{y}$
axes, which are defined by the detector needed to measure $\tilde{b}_{2,-2}$
(see Sec. \ref{sec:Tensor-observables-and-chiral-setups}). Just like
$g_{2,-2}$ is a pseudoscalar that distinguishes between molecules
with opposite handedness, $f_{2,1}$ is a pseudoscalar that distinguishes
between chiral setups of opposite handedness. Note that the detector
for $\tilde{b}_{2,-2}$ does not directly tell apart a positive $x$
from a negative $x$, or a positive $y$ from a negative $y$. It
only tells apart positive correlations of $x$ and $y$ from negative
correlations of $x$ and $y$. This is consistent with the invariance
of $f_{2,-2}$ with respect to a simultaneous inversion of $\hat{x}$
and $\hat{y}$. Furthermore, Eq. (\ref{eq:f_2m2}) shows that $f_{2,-2}$
changes sign when $\phi$ is shifted by $\pi$, and therefore $\tilde{b}_{2,-2}$
is dichroic. We remark, that although the right hand side of $f_{1,0}$
and $f_{2,-2}$ looks exactly the same after performing the vector
operations, they emerge from different setup rotational invariants
of a different physical nature: one is a scalar involving only the
$\hat{z}$ axis, and the other is a pseudoscalar involving correlation
between the $\hat{x}$ and $\hat{y}$ axes.

We consider now the simplest process leading to the enantiosensitive
but not dichroic term $\tilde{b}_{3,-2}$, i.e. we consider the absorption
of one $\omega$ photon and one $2\omega$ photon, resonantly enhanced
through a bound state as shown in Fig. \ref{fig:energy-scheme}b.
In this case we get (see Appendix \ref{subsec:appendix_blm_photoionization}) 

\begin{align}
\tilde{b}_{3,-2}\left(k\right) & =\left|A^{\left(2\right)}\right|^{2}g_{3,-2}f_{3,-2},\label{eq:b_3m2}
\end{align}

where the rotational invariant $g_{3,-2}$ is a (complex-valued) pseudoscalar\textsuperscript{\ref{fn:expressions_g}}
which makes $\tilde{b}_{3,-2}$ enantiosensitive and the setup rotational
invariant $f_{3,-2}$ is given by 

\begin{equation}
f_{3,-2}\equiv\left[\hat{y}^{\mathrm{L}}\cdot\left(\hat{z}^{\mathrm{L}}\times\hat{x}^{\mathrm{L}}\right)\right]\left(\vec{E}_{2\omega}^{\mathrm{L}*}\cdot\vec{E}_{2\omega}^{\mathrm{L}}\right)\left(\vec{E}_{\omega}^{\mathrm{L}*}\cdot\vec{E}_{\omega}^{\mathrm{L}}\right)=-E_{\omega}^{2}E_{2\omega}^{2}.\label{eq:f_3m2}
\end{equation}

$f_{3,-2}$ is also a pseudoscalar, however it does not record $\phi$
and $\tilde{b}_{3,-2}$ is therefore not dichroic. The robustness
of $f_{3,-2}$ against changes of $\phi$ means that, recording $\tilde{b}_{3,-2}$
allows distinguishing opposite enantiomers in the absence of stabilization
of the $\omega$-2$\omega$ phase shift $\phi$. 

Equations (\ref{eq:b_10})-(\ref{eq:f_3m2}) confirm our expectations
based on general symmetry arguments according to which $\tilde{b}_{1,0}$
is dichroic non-enantiosensitive, $\tilde{b}_{2,-2}$ is dichroic
and enantiosensitive, and $\tilde{b}_{3,-2}$ is enantiosensitive
non-dichroic. In addition, these equations show that $\tilde{b}_{1,0}$,
$\tilde{b}_{2,-2}$, and $\tilde{b}_{3,-2}$ are in general not zero
for the specific processes considered here. This is important because
although a given $\tilde{b}_{l,m}$ may be symmetry allowed according
to the general symmetry analysis above, further ``hidden'' symmetries\footnote{Symmetries not immediately apparent from the geometric symmetries
of the system \citep{cisneros_search_1970}.} involved in a specific process may forbid it. For example, although
the geometry of a monochromatic elliptical field allows for a non-zero
$\tilde{b}_{3,-2}$, a ``hidden'' symmetry related to the photon
ordering ensures it vanishes (see Appendix \ref{subsec:appendix_blm_photoionization}).
However, it must be kept in mind that the hidden symmetry preventing
a non-zero $\tilde{b}_{3,-2}$ for elliptical fields is specific to
the 2-photon process we investigated here, so it may be broken in
higher order processes or in the strong field regime. That is, photoionization
induced by elliptically polarized strong fields may indeed yield a
non-zero enantiosensitive non-dichroic $\tilde{b}_{3,-2}$ coefficient.
This would be analogous to how $\tilde{b}_{2,-2}$ is symmetry-forbidden
in the elliptical one-photon case (see \citep{ordonez_method_PAD})
but it is symmetry allowed in the elliptical strong-field case \citep{bashkansky_asymmetries_1988}.

\section{Conclusions\label{sec:Conclusions}}

We discussed how tensorial observables offer new opportunities for
constructing chiral setups able to distinguish between opposite enantiomers
in isotropic samples without relying on the chirality of light. As
a concrete example we considered the interaction of an $\omega$-2$\omega$
cross polarized field with isotropic samples and we found selection
rules for $l$ and $m$ that specify if a multipole coefficient $\tilde{b}_{l,m}$
is symmetry allowed, if it is dichroic (sensitive to changes of $\pi$
in the $\omega$-$2\omega$ relative phase), and if it is enantiosensitive.
These $\tilde{b}_{l,m}$ coefficients are relevant in the description
of photoelectron angular distributions and in the description of induced
multipoles of bound charge distributions (discussed elsewhere). We
found that the enantiosensitivity and the dichroism of a $\tilde{b}_{l,m}$
coefficient are in general independent of each other. That is, in
addition to the usual types of $\tilde{b}_{l,m}$ coefficients in
isotropic samples, namely: (i) non-dichroic non-enantiosensitive and
(ii) dichroic and enantiosensitive and (iii) dichroic non-enantiosensitive,
we found the more exotic possibility of (iv) enantiosensitive non-dichroic
$\tilde{b}_{l,m}$ coefficients.

We derived analytic expressions for the lowest rank $\tilde{b}_{l,m}$
coefficients in photoelectron angular distributions corresponding
to types (ii)-(iv) for the case of one- vs. two-photon absorption
and for the case of $\omega+2\omega$ absorption. Unlike type (ii)
coefficients, type (iv) coefficients allow enantiomeric discrimination
in the absence of stabilization of the $\omega$-$2\omega$ relative
phase and remain to be numerically calculated and experimentally observed.

Finally, it is also possible to obtain the type (iv) coefficient $\tilde{b}_{3,-2}$
using monochromatic light with elliptical polarization. However, hidden
symmetry prevents this coefficient in the simplest scenario of a two-photon
process. The possibility of this hidden symmetry to be violated for
higher-order or strong-field processes, or in a more refined description
of photoionization remains to be investigated. 

\FloatBarrier
\appendix
\numberwithin{equation}{section}
\numberwithin{table}{section}
\begin{section}{}

\subsection{Selection rules for an alternative orientation of the field\label{subsec:alternative_orientations}}

In Sec. \ref{sec:cross-polarized} we chose the orientation of the
field so as to simplify as much as possible the selection rules for
the $\tilde{b}_{l,m}$ coefficients. Here, for the sake of comparison,
we consider the alternative orientation\footnote{Another sensible choice, which slightly reduces the number of non-zero
$\tilde{b}_{l,m}$ coefficients, is to take the $\omega$ field along
the $z$ axis in order to obtain $\tilde{b}_{2N,m}=0$ for $m\neq0$
in the signal corresponding to $N$-photon absorption of the $\omega$
field.} used in Refs. \citep{demekhin_photoelectron_2018,demekhin_photoelectron_2019,rozen_controlling_2019},
namely

\begin{equation}
\vec{E}\left(t\right)=E_{\omega}\cos\left(\omega t\right)\hat{y}+E_{2\omega}\cos\left(2\omega t+\phi\right)\hat{x}.\label{eq:field_alt}
\end{equation}

Tables \ref{tab:conditions_alt}-\ref{tab:blm_symmetry_alt} show
the analogues of Tabs. \ref{tab:conditions}-\ref{tab:blm_symmetry}
for the orientation in Eq. \ref{eq:field_alt}. In this case, we see
that e.g. $\tilde{b}_{1,1}$, $\tilde{b}_{2,-1}$, and $\tilde{b}_{3,-2}$
are dichroic non-enantiosensitive, enantiosensitive and dichroic,
and enantiosensitive non-dichroic, respectively. These are just the
correspondingly rotated versions of the coefficients in Sec. (\ref{sec:cross-polarized}).
Note that $\tilde{b}_{2,-1}\propto\mathrm{Im}\left\{ b_{2,1}\right\} $
is indeed the coefficient discussed in Ref. \citep{demekhin_photoelectron_2019}.
\begin{center}
\par\end{center}

\begin{center}
\par\end{center}

\begin{table}
\begin{centering}
\begin{tabular}{cc}
\hline 
$\tilde{b}_{l,m}$ & $\left(l,m\right)$ condition\tabularnewline
\hline 
non-zero $2n\omega$ & (even $l-m$ and $m\geq0$) or (odd $l-m$ and $m<0$)\tabularnewline
non-zero $\left(2n+1\right)\omega$ & (odd $l-m$ and $m\geq0$) or (even $l-m$ and $m<0$)\tabularnewline
enantiosensitive & odd $l-m$\tabularnewline
dichroic & {[}(odd $l$) and ($m\geq0$){]} or {[}(even $l$) and $m<0${]}\tabularnewline
\end{tabular}
\par\end{centering}
\caption{Same as Tab. \ref{tab:conditions} for the orientation (\ref{eq:field_alt}).\label{tab:conditions_alt}}

\end{table}

\begin{table}
\begin{centering}
\begin{tabular}{ccc}
\hline 
$\tilde{b}_{l,m}$ & $\left(l,m\right)$ condition at $2n\omega$ & $(l,m)$ condition at $\left(2n+1\right)\omega$\tabularnewline
\hline 
Enantiosensitive and dichroic & (even $l$) and (odd $m<0$) & (odd $l$) and (even $m\geq0$)\tabularnewline
Enantiosensitive non-dichroic & (odd $l$) and (even $m<0$) & (even $l$) and (odd $m\geq0$)\tabularnewline
Dichroic non-enantiosensitive & (odd $l$) and (odd $m\geq0$) & (even $l$) and (even $m<0$)\tabularnewline
\hline 
\end{tabular}
\par\end{centering}
\caption{Same as Tab. \ref{tab:conditions2} for the orientation (\ref{eq:field_alt}).
\label{tab:conditions2_alt}}
\end{table}

\begin{table}
\begin{centering}
\begin{tabular}{cccccccccccccc}
\hline 
Symmetry-allowed & $\tilde{b}_{0,0}$ & $\tilde{b}_{1,1}$ & $\tilde{b}_{2,-1}$ & $\tilde{b}_{2,0}$ & $\tilde{b}_{2,2}$ & $\tilde{b}_{3,-2}$ & $\tilde{b}_{3,1}$ & $\tilde{b}_{3,3}$ & $\tilde{b}_{4,-3}$ & $\tilde{b}_{4,-1}$ & $\tilde{b}_{4,0}$ & $\tilde{b}_{4,2}$ & $\tilde{b}_{4,4}$\tabularnewline
\hline 
Enantiosensitive & N & N & Y & N & N & Y & N & N & Y & Y & N & N & N\tabularnewline
Dichroic & N & Y & Y & N & N & N & Y & Y & Y & Y & N & N & N\tabularnewline
\hline 
\end{tabular}
\par\end{centering}
\caption{Same as Tab. \ref{tab:blm_symmetry} for the orientation (\ref{eq:field_alt}).
\label{tab:blm_symmetry_alt}}

\end{table}

\subsection{Derivation of the $\boldsymbol{\tilde{b}_{l,m}}$ coefficients in
Sec. \ref{subsec:Photoionization} \label{subsec:appendix_blm_photoionization}}

\subsubsection{Derivation of $\boldsymbol{\tilde{b}_{1,0}}$ in Eq. (\ref{eq:b_10})}

\global\long\def\rotavgfour#1#2#3#4{\left[\begin{array}{c}
\left(#1\cdot#2\right)\left(#3\cdot#4\right)\\
\left(#1\cdot#3\right)\left(#2\cdot#4\right)\\
\left(#1\cdot#4\right)\left(#2\cdot#3\right)
\end{array}\right]}%

The process depicted in Fig. \ref{fig:energy-scheme}a yields a $\tilde{b}_{1,0}$
coefficient given by (see Ref. \citep{ordonez_method_PAD})

\begin{eqnarray}
\tilde{b}_{1,0} & = & \int\mathrm{d}\Omega_{k}^{\mathrm{M}}\int\mathrm{d}\varrho\tilde{Y}_{1}^{0}(\hat{k}^{\mathrm{L}})W^{\mathrm{M}}(\vec{k}^{\mathrm{M}},\varrho)\nonumber \\
 & = & \sqrt{\frac{3}{4\pi}}A^{\left(1\right)*}A^{\left(2\right)}\int\mathrm{d}\Omega_{k}^{\mathrm{M}}\int\mathrm{d}\varrho\left(\hat{k}^{\mathrm{L}}\cdot\hat{z}^{\mathrm{L}}\right)\nonumber \\
 &  & \times\left(\vec{d}_{\vec{k}^{\mathrm{M}},0}^{\mathrm{L}}\cdot\vec{E}_{2\omega}^{\mathrm{L}}\right)^{*}\left(\vec{d}_{\vec{k}^{\mathrm{M}},j}^{\mathrm{L}}\cdot\vec{E}_{\omega}^{\mathrm{L}}\right)\left(\vec{d}_{j,0}^{\mathrm{L}}\cdot\vec{E}_{\omega}^{\mathrm{L}}\right)+\mathrm{c.c.},\label{eq:b10_app}
\end{eqnarray}

where c.c. denotes the complex conjugate, $\mathrm{L}$ and $\mathrm{M}$
indicate vectors and functions in the laboratory (L) and molecular
(M) frames, $\vec{k}$ is the photoelectron momentum, $\hat{k}\equiv\vec{k}/k$,
$\varrho\equiv\alpha\beta\gamma$ is the molecular orientation specified
by the Euler angles $\alpha\beta\gamma$, $\int\mathrm{d}\varrho\equiv\int_{0}^{2\pi}\mathrm{d}\alpha\int_{0}^{\pi}\mathrm{d}\beta\int_{0}^{2\pi}\mathrm{d}\gamma\sin\beta/8\pi^{2}$
is the normalized integral over all molecular orientations, $\int\mathrm{d}\Omega_{k}^{\mathrm{M}}$
is the integral over all photoelectron directions $\hat{k}^{\mathrm{M}}$,
$W^{\mathrm{M}}(\vec{k}^{\mathrm{M}},\varrho)\equiv\vert a_{\vec{k}^{\mathrm{M}}}\left(\varrho\right)\vert^{2}$
is the photoelectron angular distribution in the molecular frame for
a particular orientation $\varrho$. $a_{\vec{k}^{\mathrm{M}}}\left(\varrho\right)=A^{\left(1\right)}(\vec{d}_{\vec{k}^{\mathrm{M}},0}^{\mathrm{L}}\cdot\vec{E}_{2\omega}^{\mathrm{L}})+A^{\left(2\right)}(\vec{d}_{\vec{k}^{\mathrm{M}},j}^{\mathrm{L}}\cdot\vec{E}_{\omega}^{\mathrm{L}})(\vec{d}_{j,0}^{\mathrm{L}}\cdot\vec{E}_{\omega}^{\mathrm{L}})$
is the probability amplitude of the state $\vert\vec{k}^{\mathrm{M}}\rangle$,
$A^{\left(1\right)}$ and $A^{\left(2\right)}$ are complex-valued
constants depending on detunings and pulse envelopes, $\vec{d}_{i,j}\equiv\langle i\vert\vec{d}\vert j\rangle$
is the dipole transition matrix element between states $\vert i\rangle$
and $\vert j\rangle$, $\vert\vec{k}\rangle$ is the scattering state
describing an outgoing plane wave with photoelectron momentum $\vec{k}$,
and $\vec{E}_{\omega}=E_{\omega}\hat{x}$ and $\vec{E}_{2\omega}=E_{2\omega}e^{-i\phi}\hat{z}$
are the Fourier amplitudes of the field (\ref{eq:field}) at frequencies
$\omega$ and $2\omega$. The vectors $\hat{k}^{\mathrm{L}}$, $\vec{d}_{i,j}^{\mathrm{L}}$,
$\vec{d}_{\vec{k}^{\mathrm{M}},j}^{\mathrm{L}}$, and $\vec{d}_{j,0}^{\mathrm{L}}$
depend on the molecular orientation $\varrho$ according to $\vec{v}^{\mathrm{L}}=S\left(\varrho\right)\vec{v}^{\mathrm{M}}$,
where $S\left(\varrho\right)$ is the rotation matrix taking vectors
from the molecular frame to the laboratory frame. Note that only the
interference between the two pathways in Fig \ref{fig:energy-scheme}
contributes to $\tilde{b}_{1,0}$ (see Sec. \ref{sec:cross-polarized}).

The integral over orientations yields (see Refs. \citep{ordonez_method_PAD}
and \citep{andrews_threedimensional_1977})

\begin{equation}
\int\mathrm{d}\varrho\left(\hat{k}^{\mathrm{L}}\cdot\hat{z}^{\mathrm{L}}\right)\left(\vec{d}_{\vec{k}^{\mathrm{M}},0}^{\mathrm{L}}\cdot\vec{E}_{2\omega}^{\mathrm{L}}\right)^{*}\left(\vec{d}_{\vec{k},j}^{\mathrm{L}}\cdot\vec{E}_{\omega}^{\mathrm{L}}\right)\left(\vec{d}_{j,0}^{\mathrm{L}}\cdot\vec{E}_{\omega}^{\mathrm{L}}\right)=\vec{g}^{\left(4\right)}\cdot M^{\left(4\right)}\vec{f}^{\left(4\right)},\label{eq:b10_orientations_integral}
\end{equation}

where $\vec{g}^{\left(4\right)}$ and $\vec{f}^{\left(4\right)}$
are vectors of molecular and setup rotational invariants, respectively, 

\begin{equation}
\vec{g}^{\left(4\right)}=\rotavgfour{\hat{k}^{\mathrm{M}}}{\vec{d}_{\vec{k}^{\mathrm{M}},0}^{\mathrm{M}*}}{\vec{d}_{\vec{k}^{\mathrm{M}},j}^{\mathrm{M}}}{\vec{d}_{j,0}^{\mathrm{M}}},\label{eq:g4}
\end{equation}

\begin{equation}
M^{\left(4\right)}=\left[\begin{array}{ccc}
4 & -1 & -1\\
-1 & 4 & -1\\
-1 & -1 & 4
\end{array}\right],
\end{equation}
 
\begin{equation}
\vec{f}^{\left(4\right)}=\rotavgfour{\hat{z}^{\mathrm{L}}}{\vec{E}_{2\omega}^{\mathrm{L}*}}{\vec{E}_{\omega}^{\mathrm{L}}}{\vec{E}_{\omega}^{\mathrm{L}}}=\left[\begin{array}{c}
\left(\hat{z}^{\mathrm{L}}\cdot\vec{E}_{2\omega}^{\mathrm{L}*}\right)\left(\vec{E}_{\omega}^{\mathrm{L}}\cdot\vec{E}_{\omega}^{\mathrm{L}}\right)\\
0\\
0
\end{array}\right].\label{eq:b10_app_f}
\end{equation}

Replacing Eqs. (\ref{eq:b10_orientations_integral})-(\ref{eq:b10_app_f})
in Eq. (\ref{eq:b10_orientations_integral}) we obtain

\begin{equation}
\tilde{b}_{1,0}=A^{\left(1\right)*}A^{\left(2\right)}g_{1,0}f_{1,0}+\mathrm{c.c.}\label{eq:b10_app-solved}
\end{equation}

where the rotational invariants are given by

\begin{align}
g_{1,0} & \equiv\frac{1}{30}\sqrt{\frac{3}{4\pi}}\int\mathrm{d}\Omega_{k}^{\mathrm{M}}\bigg[4\left(\hat{k}^{\mathrm{M}}\cdot\vec{d}_{\vec{k}^{\mathrm{M}},0}^{\mathrm{M}*}\right)\left(\vec{d}_{\vec{k}^{\mathrm{M}},j}^{\mathrm{M}}\cdot\vec{d}_{j,0}^{\mathrm{M}}\right)\nonumber \\
 & -\left(\hat{k}^{\mathrm{M}}\cdot\vec{d}_{\vec{k}^{\mathrm{M}},j}^{\mathrm{M}}\right)\left(\vec{d}_{\vec{k}^{\mathrm{M}},0}^{\mathrm{M}*}\cdot\vec{d}_{j,0}^{\mathrm{M}}\right)-\left(\hat{k}^{\mathrm{M}}\cdot\vec{d}_{j,0}^{\mathrm{M}}\right)\left(\vec{d}_{\vec{k}^{\mathrm{M}},0}^{\mathrm{M}*}\cdot\vec{d}_{\vec{k}^{\mathrm{M}},j}^{\mathrm{M}}\right)\bigg],\label{eq:g10_app}
\end{align}

\begin{align}
f_{1,0} & \equiv\left(\hat{z}^{\mathrm{L}}\cdot\vec{E}_{2\omega}^{\mathrm{L}*}\right)\left(\vec{E}_{\omega}^{\mathrm{L}}\cdot\vec{E}_{\omega}^{\mathrm{L}}\right)\label{eq:f10_app_aux}
\end{align}

We remark that although Eq. (\ref{eq:f10_app_aux}) may suggest that
Eqs. (\ref{eq:b10_app-solved})-(\ref{eq:f10_app_aux}) are valid
for arbitrary $\vec{E}_{\omega}$ and $\vec{E}_{2\omega}$, they are
not. We have kept the general vectorial form of the rotational invariant
$(\hat{z}^{\mathrm{L}}\cdot\vec{E}_{2\omega}^{\mathrm{L}*})(\vec{E}_{\omega}^{\mathrm{L}}\cdot\vec{E}_{\omega}^{\mathrm{L}})$
for illustrative purposes only (see discussion in Sec. \ref{subsec:Photoionization}).
As can be seen from Eq. (\ref{eq:b10_app_f}), in deriving Eqs. (\ref{eq:b10_app-solved})-(\ref{eq:f10_app_aux})
we have already taken into account that $\vec{E}_{2\omega}^{\mathrm{L}}\parallel\hat{z}^{\mathrm{L}}$
and $\vec{E}_{\omega}^{\mathrm{L}}\parallel\hat{x}^{\mathrm{L}}$.

\subsubsection{Derivation of $\boldsymbol{\tilde{b}_{2,-2}}$ in Eq. (\ref{eq:b_2m2})}

Similarly, for the $\tilde{b}_{2,-2}$ coefficient we have \citep{ordonez_method_PAD}

\begin{eqnarray}
\tilde{b}_{2,-2} & = & \int\mathrm{d}\Omega_{k}^{\mathrm{M}}\int\mathrm{d}\varrho\tilde{Y}_{2}^{-2}(\vec{k}^{\mathrm{L}})W^{\mathrm{M}}(\vec{k}^{\mathrm{M}},\varrho)\nonumber \\
 & = & \frac{1}{2}\sqrt{\frac{15}{\pi}}A^{\left(1\right)*}A^{\left(2\right)}\int\mathrm{d}\Omega_{k}^{\mathrm{M}}\int\mathrm{d}\varrho\left(\hat{k}^{\mathrm{L}}\cdot\hat{x}^{\mathrm{L}}\right)\left(\hat{k}^{\mathrm{L}}\cdot\hat{y}^{\mathrm{L}}\right)\nonumber \\
 &  & \times\left(\vec{d}_{\vec{k}^{\mathrm{M}},0}^{\mathrm{L}}\cdot\vec{E}_{2\omega}^{\mathrm{L}}\right)^{*}\left(\vec{d}_{\vec{k}^{\mathrm{M}},j}^{\mathrm{L}}\cdot\vec{E}_{\omega}^{\mathrm{L}}\right)\left(\vec{d}_{j,0}^{\mathrm{L}}\cdot\vec{E}_{\omega}^{\mathrm{L}}\right)+\mathrm{c.c.}\label{eq:b2m2_app}\\
 & = & A^{\left(1\right)*}A^{\left(2\right)}g_{2,-2}f_{2,-2}+\mathrm{c.c.}\label{eq:b2m2_app_solved}
\end{eqnarray}

where the integral over orientations was performed analogously to
what we did for $\tilde{b}_{1,0}$ \citep{andrews_threedimensional_1977},
and the rotational invariants are given by 

\begin{equation}
g_{2,-2}\equiv\frac{1}{4\sqrt{15\pi}}\int\mathrm{d}\Omega_{k}^{\mathrm{M}}\bigg\{\left[\hat{k}^{\mathrm{M}}\cdot\left(\vec{d}_{\vec{k}^{\mathrm{M}},0}^{\mathrm{M}*}\times\vec{d}_{\vec{k}^{\mathrm{M}},j}^{\mathrm{M}}\right)\right]\left(\hat{k}^{\mathrm{M}}\cdot\vec{d}_{j,0}^{\mathrm{M}}\right)+\left[\hat{k}^{\mathrm{M}}\cdot\left(\vec{d}_{\vec{k}^{\mathrm{M}},0}^{\mathrm{M}*}\times\vec{d}_{j,0}^{\mathrm{M}}\right)\right]\left(\hat{k}^{\mathrm{M}}\cdot\vec{d}_{\vec{k}^{\mathrm{M}},j}^{\mathrm{M}}\right)\bigg\},\label{eq:g2m2}
\end{equation}

\begin{align}
f_{2,-2} & \equiv\left[\hat{x}^{\mathrm{L}}\cdot\left(\hat{y}^{\mathrm{L}}\times\vec{E}_{2\omega}^{\mathrm{L}*}\right)\right]\left(\vec{E}_{\omega}^{\mathrm{L}}\cdot\vec{E}_{\omega}^{\mathrm{L}}\right)\label{eq:f2m2_app_aux}
\end{align}

Like in the case of $\tilde{b}_{1,0}$, we remark that despite the
general aspect of Eq. (\ref{eq:f2m2_app_aux}), Eqs. (\ref{eq:b2m2_app_solved})-(\ref{eq:f2m2_app_aux})
are valid specifically for $\vec{E}_{2\omega}^{\mathrm{L}}\parallel\hat{z}^{\mathrm{L}}$
and $\vec{E}_{\omega}^{\mathrm{L}}\parallel\hat{x}^{\mathrm{L}}$.
Furthermore, we point out that when dealing with integrals over orientations
involving five or more scalar products, the number of rotational invariants
that can be formed is such that they are no longer linearly independent
from each other \citep{andrews_threedimensional_1977}. As a result
it is possible to write the result of the integral in several different
ways that, although perfectly equivalent, are not evidently related
to each other at first sight. For example, by changing the order of
the scalar products in such a way that $(\vec{d}_{\vec{k},j}^{\mathrm{L}}\cdot\vec{E}_{\omega}^{\mathrm{L}})$
exchanges its position with $(\hat{k}^{\mathrm{L}}\cdot\hat{x}^{\mathrm{L}})$
in Eq. (\ref{eq:b2m2_app_solved}) one obtains

\begin{equation}
\tilde{b}_{2,-2}=A^{\left(1\right)*}A^{\left(2\right)}g_{2,-2}^{\prime}f_{2,-2}^{\prime}+\mathrm{c.c.},
\end{equation}

where

\begin{align}
g_{2,-2}^{\prime} & \equiv\frac{1}{4\sqrt{15\pi}}\int\mathrm{d}\Omega_{k}^{\mathrm{M}}\bigg\{2\left[\vec{d}_{\vec{k}^{\mathrm{M}},j}^{\mathrm{M}}\cdot\left(\hat{k}^{\mathrm{M}}\times\vec{d}_{\vec{k}^{\mathrm{M}},0}^{\mathrm{M}*}\right)\right]\left(\hat{k}^{\mathrm{M}}\cdot\vec{d}_{j,0}^{\mathrm{M}}\right)\nonumber \\
 & -\left[\vec{d}_{\vec{k}^{\mathrm{M}},j}^{\mathrm{M}}\cdot\left(\hat{k}^{\mathrm{M}}\times\vec{d}_{j,0}^{\mathrm{M}}\right)\right]\left(\vec{d}_{\vec{k}^{\mathrm{M}},0}^{\mathrm{M}*}\cdot\hat{k}^{\mathrm{M}}\right)+\left[\vec{d}_{\vec{k}^{\mathrm{M}},j}^{\mathrm{M}}\cdot\left(\vec{d}_{\vec{k}^{\mathrm{M}},0}^{\mathrm{M}*}\times\vec{d}_{j,0}^{\mathrm{M}}\right)\right]\left(\hat{k}^{\mathrm{M}}\cdot\hat{k}^{\mathrm{M}}\right)\bigg\},
\end{align}

\begin{equation}
f_{2,-2}^{\prime}\equiv\left[\vec{E}_{\omega}^{\mathrm{L}}\cdot\left(\hat{y}^{\mathrm{L}}\times\vec{E}_{2\omega}^{\mathrm{L}*}\right)\right]\left(\hat{x}^{\mathrm{L}}\cdot\vec{E}_{\omega}^{\mathrm{L}}\right),
\end{equation}
and again we assumed that $\vec{E}_{2\omega}^{\mathrm{L}}\parallel\hat{z}^{\mathrm{L}}$
and $\vec{E}_{\omega}^{\mathrm{L}}\parallel\hat{x}^{\mathrm{L}}$.
Writing the explicit expressions $\vec{E}_{\omega}=E_{\omega}\hat{x}$
and $\vec{E}_{2\omega}^{*}=E_{2\omega}e^{i\phi}\hat{z}$ one can show
that $f_{2,-2}^{\prime}=f_{2,-2}$. And using standard vectorial algebra
relations one can show that $g_{2,-2}^{\prime}=g_{2,2}$. The latter
equality reflects the fact that, for four arbitrary vectors $\vec{a}$,
$\vec{b}$, $\vec{c}$, $\vec{d}$, the \emph{three} different rotational
invariants $r_{1}\equiv[\vec{a}\cdot(\vec{b}\times\vec{c})](\vec{b}\cdot\vec{d})$,
$r_{2}\equiv[\vec{a}\cdot(\vec{b}\times\vec{d})][\vec{c}\cdot\vec{b}]$,
and $r_{3}\equiv[\vec{a}\cdot(\vec{c}\times\vec{d})](\vec{b}\cdot\vec{b})$
can be written as a linear combination of the \emph{two} rotational
invariants $r_{1}$ and $r_{4}\equiv[\vec{b}\cdot(\vec{c}\times\vec{d})](\vec{b}\cdot\vec{a})$.
Care must therefore be taken when looking for interpretations that
depend on the particular ordering of the vectors appearing in the
rotational invariants. 

\subsubsection{Derivation of $\boldsymbol{\tilde{b}_{3,-2}}$ for $\boldsymbol{\omega+2\omega}$
{[}Eq. (\ref{eq:b_3m2}){]}}

The process depicted in Fig. \ref{fig:energy-scheme}b yields a $\tilde{b}_{3,-2}$
coefficient given by \citep{ordonez_method_PAD}

\begin{align}
\tilde{b}_{3,-2}\left(k\right) & =\int\mathrm{d}\Omega_{k}^{\mathrm{M}}\int\mathrm{d}\varrho\,\tilde{Y}_{3}^{-2}(\hat{k}^{\mathrm{L}})W^{\mathrm{M}}(\vec{k}^{\mathrm{M}},\varrho)\nonumber \\
 & =\left|A^{\left(2\right)}\right|^{2}\frac{1}{2}\sqrt{\frac{105}{\pi}}\int\mathrm{d}\Omega_{k}^{\mathrm{M}}\int\mathrm{d}\varrho\,\left(\hat{k}^{\mathrm{L}}\cdot\hat{x}^{\mathrm{L}}\right)\left(\hat{k}^{\mathrm{L}}\cdot\hat{y}^{\mathrm{L}}\right)\left(\hat{k}^{\mathrm{L}}\cdot\hat{z}^{\mathrm{L}}\right)\nonumber \\
 & \times\left|\vec{d}_{\vec{k}^{\mathrm{M}},j}^{\mathrm{L}}\cdot\vec{E}_{2\omega}^{\mathrm{L}}\right|^{2}\left|\vec{d}_{j,0}^{\mathrm{L}}\cdot\vec{E}_{\omega}^{\mathrm{L}}\right|^{2}\label{eq:b3m2_app}\\
 & =\left|A^{\left(2\right)}\right|^{2}g_{3,-2}f_{3,-2}\label{eq:b3m2_app_solved}
\end{align}

where the integral over orientations was performed analogously to
what we did for $\tilde{b}_{1,0}$\footnote{This is considerably simplified by ordering the scalar products in
the orientation integral as $(\hat{k}^{\mathrm{L}}\cdot\hat{x}^{\mathrm{L}})(\hat{k}^{\mathrm{L}}\cdot\hat{z}^{\mathrm{L}})(\vec{d}_{j,0}^{\mathrm{L}*}\cdot\vec{E}_{\omega}^{\mathrm{L}*})(\hat{k}^{\mathrm{L}}\cdot\hat{y}^{\mathrm{L}})(\vec{d}_{\vec{k}^{\mathrm{M}},j}^{\mathrm{L}*}\cdot\vec{E}_{2\omega}^{\mathrm{L}*})(\vec{d}_{\vec{k}^{\mathrm{M}},j}^{\mathrm{L}}\cdot\vec{E}_{2\omega}^{\mathrm{L}})(\vec{d}_{j,0}^{\mathrm{L}}\cdot\vec{E}_{\omega}^{\mathrm{L}})$
and using table III in Ref. \citep{andrews_threedimensional_1977}.}, the rotational invariants are given by

\begin{equation}
g_{3,-2}\equiv\frac{1}{8\sqrt{105\pi}}\int\mathrm{d}\Omega_{k}^{\mathrm{M}}[\hat{k}^{\mathrm{M}}\cdot(\vec{d}_{j,0}^{\mathrm{M}}\times\vec{d}_{\vec{k}^{\mathrm{M}},j}^{\mathrm{M}*})]\left[(\vec{d}_{\vec{k},j}^{\mathrm{M}}\cdot\vec{d}_{j,0}^{\mathrm{M}})-5(\hat{k}^{\mathrm{M}}\cdot\vec{d}_{\vec{k}^{\mathrm{M}},j}^{\mathrm{M}})(\hat{k}^{\mathrm{M}}\cdot\vec{d}_{j,0}^{\mathrm{M}})\right]+\mathrm{c.c.},
\end{equation}

\begin{equation}
f_{3,-2}\equiv\left[\hat{y}^{\mathrm{L}}\cdot\left(\hat{z}^{\mathrm{L}}\times\hat{x}^{\mathrm{L}}\right)\right]\left(\vec{E}_{2\omega}^{\mathrm{L}*}\cdot\vec{E}_{2\omega}^{\mathrm{L}}\right)\left(\vec{E}_{\omega}^{\mathrm{L}*}\cdot\vec{E}_{\omega}^{\mathrm{L}}\right),
\end{equation}

and we relied on $\vec{E}_{2\omega}^{\mathrm{L}}\parallel\hat{z}^{\mathrm{L}}$
and $\vec{E}_{\omega}^{\mathrm{L}}\parallel\hat{x}^{\mathrm{L}}$.

\subsubsection{Derivation of $\boldsymbol{\tilde{b}_{3,-2}}$ for $\boldsymbol{2\omega+\omega}$:
symmetry in photon ordering}

Although we could calculate the expression for $\tilde{b}_{3,-2}$
for the photon ordering $2\omega+\omega$ analogously to how we did
it for the opposite photon ordering, the great number of different
rotational invariants for this case would obscure the relation between
$\tilde{b}_{3,-2}$ in the two cases. Here we follow a more instructive
and powerful approach that relies on the symmetry of the structure
of the multiphoton amplitudes (see also the derivation of $\tilde{b}_{2,-2}$
in one-photon ionization in Ref. \citep{ordonez_method_PAD}).

To see how the photon ordering affects the value of $\tilde{b}_{3,-2}$
let us first define the function

\begin{align}
I(\hat{a}^{\mathrm{L}},\hat{b}^{\mathrm{L}},\hat{c}^{\mathrm{L}}) & \equiv E_{\omega}^{2}E_{2\omega}^{2}\int\mathrm{d}\varrho\,\left(\hat{k}^{\mathrm{L}}\cdot\hat{a}^{\mathrm{L}}\right)\left(\hat{k}^{\mathrm{L}}\cdot\hat{b}^{\mathrm{L}}\right)\left(\hat{k}^{\mathrm{L}}\cdot\hat{c}^{\mathrm{L}}\right)\left|\vec{d}_{\vec{k}^{\mathrm{M}},j}^{\mathrm{L}}\cdot\hat{c}^{\mathrm{L}}\right|^{2}\left|\vec{d}_{j,0}^{\mathrm{L}}\cdot\hat{a}^{\mathrm{L}}\right|^{2}.\label{eq:I_abc}
\end{align}

The integral $I_{\omega,2\omega}$ appearing in Eq. (\ref{eq:b3m2_app})
and corresponding to the absorption of $\omega$ followed by absorption
of $2\omega$ (see Fig. \ref{fig:energy-scheme}b) reads as

\begin{align}
I_{\omega,2\omega} & \equiv\int\mathrm{d}\varrho\,\left(\hat{k}^{\mathrm{L}}\cdot\hat{x}^{\mathrm{L}}\right)\left(\hat{k}^{\mathrm{L}}\cdot\hat{y}^{\mathrm{L}}\right)\left(\hat{k}^{\mathrm{L}}\cdot\hat{z}^{\mathrm{L}}\right)\left|\vec{d}_{\vec{k}^{\mathrm{M}},j}^{\mathrm{L}}\cdot\vec{E}_{2\omega}^{\mathrm{L}}\right|^{2}\left|\vec{d}_{j,0}^{\mathrm{L}}\cdot\vec{E}_{\omega}^{\mathrm{L}}\right|^{2}\nonumber \\
 & =I(\hat{x}^{\mathrm{L}},\hat{y}^{\mathrm{L}},\hat{z}^{\mathrm{L}}),
\end{align}

where we used $\vec{E}_{\omega}=E_{\omega}\hat{x}$ and $\vec{E}_{2\omega}=E_{2\omega}e^{-i\phi}\hat{z}$.
In contrast, the integral $I_{2\omega,\omega}$ corresponding to the
opposite photon ordering (absorption of $2\omega$ followed by absorption
of $\omega$) reads as 

\begin{align}
I_{2\omega,\omega} & \equiv\int\mathrm{d}\varrho\,\left(\hat{k}^{\mathrm{L}}\cdot\hat{x}^{\mathrm{L}}\right)\left(\hat{k}^{\mathrm{L}}\cdot\hat{y}^{\mathrm{L}}\right)\left(\hat{k}^{\mathrm{L}}\cdot\hat{z}^{\mathrm{L}}\right)\left|\vec{d}_{\vec{k}^{\mathrm{M}},j}^{\mathrm{L}}\cdot\vec{E}_{\omega}^{\mathrm{L}}\right|^{2}\left|\vec{d}_{j,0}^{\mathrm{L}}\cdot\vec{E}_{2\omega}^{\mathrm{L}}\right|^{2}\nonumber \\
 & =I\left(\hat{z}^{\mathrm{L}},\hat{y}^{\mathrm{L}},\hat{x}^{\mathrm{L}}\right)\nonumber \\
 & =-I(-\hat{z}^{\mathrm{L}},\hat{y}^{\mathrm{L}},\hat{x}^{\mathrm{L}}).
\end{align}
Now we define a second laboratory frame $\{\hat{x}^{\mathrm{L}^{\prime}},\hat{y}^{\mathrm{L}^{\prime}},\hat{z}^{\mathrm{L}^{\prime}}\}$
rotated with respect to $\{\hat{x}^{\mathrm{L}},\hat{y}^{\mathrm{L}},\hat{z}^{\mathrm{L}}\}$
by $\pi/2$ around $\hat{y}^{\mathrm{L}}$, such that it satisfies
$\{\hat{x}^{\mathrm{L}^{\prime}},\hat{y}^{\mathrm{L}^{\prime}},\hat{z}^{\mathrm{L}^{\prime}}\}=\{-\hat{z}^{\mathrm{L}},\hat{y}^{\mathrm{L}},\hat{x}^{\mathrm{L}}\}$.
Physically, this corresponds to a rigid rotation of the experimental
setup as a whole (i.e. field and detectors). Using this rotated frame
and taking into account that $I(\hat{x}^{\mathrm{L}},\hat{y}^{\mathrm{L}},\hat{z}^{\mathrm{L}})$
depends only on rotational invariants {[}and therefore $I(\hat{x}^{\mathrm{L}^{\prime}},\hat{y}^{\mathrm{L}^{\prime}},\hat{z}^{\mathrm{L}^{\prime}})=I(\hat{x}^{\mathrm{L}},\hat{y}^{\mathrm{L}},\hat{z}^{\mathrm{L}})${]},
we obtain

\begin{align}
I_{2\omega,\omega} & =-I(\hat{x}^{\mathrm{L}^{\prime}},\hat{y}^{\mathrm{L}^{\prime}},\hat{z}^{\mathrm{L}^{\prime}})\nonumber \\
 & =-I_{\omega,2\omega}.
\end{align}

This means that the expression for $\tilde{b}_{3,-2}$ in the photon
ordering $2\omega+\omega$ is exactly the same as for the photon ordering
$\omega+2\omega$ up to a minus sign. However, since for a fixed molecular
spectrum the coupling coefficient $A^{\left(2\right)}$ strongly depends
on detunings (and therefore photon ordering), if one of the photon
orderings is resonant it will dominate. Furthermore, in practice each
photon ordering might be resonant with different transitions, and
one should therefore use different transition dipole matrix elements
for each ordering. 

Since the derivation we just presented is actually independent of
the frequencies, we have that, provided $\omega_{1}$ and $\omega_{2}$
have linear polarizations perpendicular to each other, then 

\begin{equation}
I_{\omega_{1},\omega_{2}}=-I_{\omega_{2},\omega_{1}},\label{eq:hidden_symmetry}
\end{equation}

and we can therefore conclude that $\tilde{b}_{3,-2}\rightarrow0$
as $\omega_{1}\rightarrow\omega_{2}$. Note that this is analogous
to the corresponding result for the polarization in sum-frequency
generation \citep{giordmaine_nonlinear_1965}.

\subsubsection{Vanishing of $\tilde{\boldsymbol{b}}_{3,-2}$ for elliptical light}

Using relation (\ref{eq:hidden_symmetry}) it is simple to establish
why, although a monochromatic elliptical field has a geometry that
allows for a non-zero $\tilde{b}_{3,-2}$ in an $\omega+2\omega$
process, the ``hidden'' symmetry in Eq. (\ref{eq:hidden_symmetry})
forces it to vanish. To see this, note that for an elliptical field
we have 

\begin{align}
\vec{E}\left(t\right) & =\vec{E}_{\omega}e^{-i\omega t}+\mathrm{c.c.}
\end{align}

with $\vec{E}_{\omega}=E_{x}\hat{x}+iE_{y}\hat{y}$ and therefore
{[}c.f. Eq. (\ref{eq:b3m2_app}){]}

\begin{align}
W^{\mathrm{M}}(\vec{k}^{\mathrm{M}},\varrho) & =\left|A^{\left(2\right)}\right|^{2}\left|\vec{d}_{\vec{k}^{\mathrm{M}},j}^{\mathrm{L}}\cdot\vec{E}_{\omega}^{\mathrm{L}}\right|^{2}\left|\vec{d}_{j,0}^{\mathrm{L}}\cdot\vec{E}_{\omega}^{\mathrm{L}}\right|^{2}\nonumber \\
 & =\left|A^{\left(2\right)}\right|^{2}\bigg\{\left|\vec{d}_{\vec{k}^{\mathrm{M}},j}^{\mathrm{L}}\cdot E_{x}\hat{x}^{\mathrm{L}}\right|^{2}\left|\vec{d}_{j,0}^{\mathrm{L}}\cdot E_{x}\hat{x}^{\mathrm{L}}\right|^{2}+\left|\vec{d}_{\vec{k}^{\mathrm{M}},j}^{\mathrm{L}}\cdot E_{y}\hat{y}^{\mathrm{L}}\right|^{2}\left|\vec{d}_{j,0}^{\mathrm{L}}\cdot E_{y}\hat{y}^{\mathrm{L}}\right|^{2}\nonumber \\
 & +\left|\vec{d}_{\vec{k}^{\mathrm{M}},j}^{\mathrm{L}}\cdot E_{x}\hat{x}^{\mathrm{L}}\right|^{2}\left|\vec{d}_{j,0}^{\mathrm{L}}\cdot E_{y}\hat{y}^{\mathrm{L}}\right|^{2}+\left|\vec{d}_{\vec{k}^{\mathrm{M}},j}^{\mathrm{L}}\cdot E_{y}\hat{y}^{\mathrm{L}}\right|^{2}\left|\vec{d}_{j,0}^{\mathrm{L}}\cdot E_{x}\hat{x}^{\mathrm{L}}\right|^{2}\bigg\},
\end{align}
that is, we can decompose the process into four pathways. Two of them
involve either two $x$-polarized photons or two $y$-polarized photons
and the associated geometrical symmetry prevents them from contributing
to $\tilde{b}_{3,-2}$. The other two terms involve absorption of
one $x$-polarized photon and one $y$-polarized photon and their
lower geometrical symmetry allows for contributions to $\tilde{b}_{3,-2}$.
However, since these two terms correspond to opposite photon orderings
satisfying $\omega_{1}\rightarrow\omega_{2}$, the photon-ordering
symmetry in Eq. (\ref{eq:hidden_symmetry}) implies that their contributions
will cancel each other exactly.

\end{section}

\bibliographystyle{apsrev4-1}
\bibliography{MyLibrary}

\end{document}